# Piezomagnetic Switching of Nonvolatile Antiferromagnetic States

Xilai Bao[1,2†], Oleksandr V. Pylypovskyi[3,4†], Huali Yang[1*], Yali Xie[1*], Damien Faurie[5], Fatih Zighem[5], Sophie F. Weber[6,7], Jiabin Wang[8], Jiachen Liang[1], Hong Xu[1,8], Ruoan Zou[1,2], Huatao Jiang[1,2], Dong Han[1], Pavlo Makushko[3], Xiaotao Wang[3], Lin Guo[3], Proloy T. Das[3], Nicola A. Spaldin[6], Denys Makarov[3*], and Run-Wei Li[1,8*]

*[1]Zhejiang Province Key Laboratory of Magnetic Materials and Application Technology, Ningbo Institute of Materials Technology and Engineering, Chinese Academy of Sciences, Ningbo 315201, P. R. China*

*[2]School of Future Technology, University of Chinese Academy of Sciences, Beijing 100049, P. R. China*

*[3]Helmholtz-Zentrum Dresden-Rossendorf e.V., Institute of Ion Beam Physics and Materials Research, 01328 Dresden, Germany*

*[4]Kyiv Academic University, Kyiv 03142, Ukraine*

*[5]LSPM-CNRS UPR3407, Université Sorbonne Paris Nord, 93430 Villetaneuse, France*

*[6]Materials Theory, ETH Zürich, Wolfgang-Pauli-Strasse 27, 8093 Zürich, Switzerland*

*[7]Chalmers University of Technology, 412 96 Gothenburg, Sweden*

*[8]Eastern Institute of Technology, Ningbo 315200, P. R. China*

† These authors contributed equally to this work.

* Corresponding authors: yanghl@nimte.ac.cn (Huali Yang), xieyl@nimte.ac.cn (Yali Xie), d.makarov@hzdr.de (Denys Makarov), rwli@eitech.edu.cn (Run-Wei Li)

## Abstract

Prospective spintronic memory and logic devices will benefit from the negligible stray field and ultrafast magnetic dynamics inherent to antiferromagnets[1]. However, realizing isothermal, non-volatile, and deterministic switching of antiferromagnetic states remains a key challenge[2,3]. Here, we propose a piezomagnetic writing scheme in triangular $Mn_3Ir$-based memory cells, with readout achieved via the exchange bias effect. Our approach enables deterministic and nonvolatile switching of the antiferromagnetic states, which exhibit exceptional robustness against external perturbations. The switching mechanism is ascribed to piezomagnetic effect of $Mn_3Ir$ combined with the interfacial Dzyaloshinskii-Moriya interaction at the antiferromagnet-ferromagnet interface. This scheme overcomes the speed limitations imposed by conventional isothermal methods based on isothermal crystallization mechanism[4]. Our findings highlight the potential of piezomagnetic effects in designing advanced spintronic devices, providing an efficient pathway for manipulating antiferromagnetic states and developing energy-efficient memory technology.

## Introduction

Compared to ferromagnets (FM), antiferromagnetic (AF) materials exhibit unique advantages including ultrafast dynamics, robustness against external fields and absence of stray fields, making them appealing candidates for high-density information storage and THz spintronic devices[5-7]. Noncollinear antiferromagnets, in particular, can exhibit ferromagnet-like transport and optical phenomena, such as anomalous Hall effect (AHE)[8-10] and magneto-optical responses[11,12], owing to their complex spin textures and reduced crystalline symmetries[1]. Thus, noncollinear antiferromagnets provide a compelling substitute for ferromagnets in prospective high-performance spintronic and spin-orbitronic devices[13,14]. To enable a reliable writing functionality as required for an AF memory, it is imperative to achieve deterministic control of the magnetic state of noncollinear antiferromagnets. Beyond the early success of current-induced control in collinear CuMnAs[15], extensive efforts have also been dedicated to the deterministic manipulation of strongly noncollinear antiferromagnets. This can be achieved using spin-orbit torque (SOT) generated in adjacent heavy metals (HM)[9,16-18]. Based on these advances, all-antiferromagnetic tunnel junctions featuring information writing via SOT or spin transfer torque have been successfully fabricated[13,19,20]. As the typically utilized current-driven approaches involve substantial Joule heating potentially impacting the device reliability, there is an ongoing development of isothermal approaches to manipulate AF states, including magnetoelectrical effect[21-23], ionic liquid[24,25], strain[2,26-28] and isothermal crystallization[4].

Strain has been identified as an efficient tool to alter the AF states isothermally, featured by their low energy consumption and minimized Joule heating[28-30]. The application of strain modulates the magnetic anisotropy via the magnetostrictive effect[30,31], adjusts the magnetic transition temperature and can enhance multiferroic behavior[32,33]. Among these magnetoelastic effects, the piezomagnetic effect is characterized by a linear magneto-mechanical coupling and is commonly observed in a variety of noncollinear antiferromagnets, including $Mn_3X$ (X = Ga, Sn, Pt, Ir)[2,26,32,34,35] and antiperovskite materials ($Mn_3NiN$)[36]. These properties make piezomagnetism a beneficial approach for developing ultrafast and energy-efficient AF spintronics[27,37]. However, nonvolatile control of AF states through strain has yet to be achieved, despite its significance for applications in AF memory and logic devices. The challenge stems from the fact that the manipulation of the direction of spins is closely associated with elastic deformations of the crystal lattice[38], which indicates that the magnetic state returns to its original configuration when the strain is released[3,39].

Among the above antiferromagnets, $Mn_3Ir$ stands out due to its exceptional thermal stability and proven technological relevance for spintronics[40,41]. $Mn_3Ir$ is a triangular antiferromagnet characterized by the space group $Pm\bar{3}m$ (No. 221 with the crystallographic point group $m\bar{3}m$) and magnetic point group $\bar{3}m'$. This noncollinear spin structure[4,40,42] and strong spin-orbit coupling[12,43,44] of $Mn_3Ir$ enable an efficient spin manipulation through a large magnetoelastic effect. Here, we demonstrate deterministic writing of binary information in an AF memory cell based on polycrystalline antiferromagnetic $Mn_3Ir$ thin films (Fig. 1a). For a concise description of magnetic states of $Mn_3Ir$, we consider three vectors of magnetic moments within the unit cell, focusing on the compensated yet chiral arrangement (Fig. 1b,d). The writing in AF bits is facilitated by breaking the energy symmetry between the two opposite AF domains through a tensile strain applied to $Mn_3Ir$ (Fig. 1c,e), while the readout is enabled via the exchange bias (EB) effect provided by an adjacent FM layer of $[Co/Pt]_3$. We assign the AF state "1" to the case when the magnetic hysteresis loop exhibits a positive perpendicular exchange bias field ($H_{PEB}$, Fig. 1f), and AF state "0" for a negative $H_{PEB}$ (Fig. 1g). We first demonstrate the continuous "write" and "read" behaviors using the above strategy through the measurement of the remanence values assessed via anomalous Hall measurements (Supporting Note 6). The reproducibility of this switching process is depicted in Fig. 1h, where we see that two AF states could be set repeatedly under mechanical strain and cyclic magnetic field. Our analytical and density functional theory (DFT) calculations reveal that the manipulation of AF states is due to the piezomagnetic effect in $Mn_3Ir$, assisted by interfacial Dzyaloshinskii-Moriya interaction (iDMI) at the AF/FM interface, which significantly reduces the energy barrier for the AF domain reversal. Moreover, owing to the pinning effect of the AF domains, the written AF states exhibit nonvolatility as well as exceptional robustness against external magnetic field and mechanical strain. Our results demonstrate a novel approach for nonvolatile and precise manipulation of AF states in thin-film antiferromagnets, which is crucial for the fabrication of magnetoelectronic devices including energy efficient nonvolatile memories.

## Results

### Structure analysis and strain characterization

$Mn_3Ir/[Co/Pt]_3$ stacks were deposited on flexible 50-$\mu$m-thick polyimide foils using magnetron sputtering at room temperature. High-resolution cross-sectional transmission electron microscopy (TEM) images reveal a (1 1 1) texture and a chemically disordered $\gamma$-$Mn_3Ir$ phase[4,31], which are further verified via small angle X-ray scattering (SAXS) spectrum collected in the OOP geometry (Supporting Note 1). Additionally, the IP SAXS spectrum indicates that the $Mn_3Ir$ layer is polycrystalline and there is no evidence that further crystallization happens in the film (Supporting Note 1). In this respect, our samples are distinct from the spontaneous exchange bias (SEB) effect through isothermal crystallization[4], where the as-deposited amorphous $Mn_3Ir$ layer transitioned into the crystallized phase after many hours. Consequently, the manipulation of the AF state in our case is not related to the isothermal crystallization of the $Mn_3Ir$ layer.

To evaluate the impact of the tensile strain on the crystal structure of $Mn_3Ir$, OOP SAXS spectra were captured under different strains in situ and then averaged into 1D scattering spectra. A correlation between the scattering intensity and scattering angle under different tensile strains is depicted in Supporting Note 2. We observe a reversible shift of the SAXS peak upon the application and release of tensile strain to the polyimide (maximal $\varepsilon_{PI}$ = 3%), indicating elastic deformations in the metal layer stack. This elastic range was further verified by the surface morphology captured using scanning electron microscopy (SEM) (Supporting Note 2). Due to an incomplete strain transfer from the thin polymeric substrate to the metal layer stack[31], the relative variation of the crystal parameters of $Mn_3Ir$ along the stretching direction, $\varepsilon$, is smaller than $\varepsilon_{PI}$. In particular, $\varepsilon_{PI}$ of 3% corresponds to $\varepsilon$ of 1.2% (Supporting Note 2). Hereafter, we use $\varepsilon$ as a representative parameter for the strain experienced by the metal layer stack.

**Deterministic manipulation of nonvolatile antiferromagnetic states via strain**

In our experiments, $Mn_3Ir/[Co/Pt]_3$ stacks were fabricated without applying any external magnetic field. The magnetic hysteresis loops of the as-deposited samples show a prominent perpendicular magnetic anisotropy and no loop shift (Supporting Note 3). The zero $H_{PEB}$ indicates an equal distribution of AF domains within the $Mn_3Ir$ layer. By stretching the sample at room temperature, with the $[Co/Pt]_3$ multilayers magnetized in an OOP magnetic field of −2 kOe (the detailed process is in Supporting Note 4), $H_{PEB}$ develops instantly (Fig. 2a, upper panel). With increasing the tensile strain, the $H_{PEB}$ develops from zero and increases almost linearly to approximately 120 Oe at $\varepsilon$ = 1.2% (Fig. 2c). The emergence of a non-zero $H_{PEB}$ indicates the formation of a preferential Néel-vector configuration in $Mn_3Ir$, which is also observed by using MOKE microscopy (Supporting Note 5), AHE characterization (Supporting Note 6), and magnetometric hysteresis loop measurements (Supporting Note 7).

Interestingly, the written AF state show nonvolatility upon release of the mechanical strain. As is shown in Fig. 2a (lower panel) and Fig. 2c, upon the removal of strain, $H_{PEB}$ remains unchanged. Noticeably, the nonvolatility of $H_{PEB}$ is intriguing in consideration of the relaxed lattice structure in the unstrained sample (Supporting Note 2). Upon strain release, the recovery of the lattice structure should restore the bulk AF structures to their equilibrium state, thus eliminating preferential AF states and resulting in zero $H_{PEB}$. However, our results indicate that the stretching combined with the magnetized FM layer leads to the selection of a dominant AF domain that keeps its stability upon release.

Our approach to manipulating the two AF states (Fig. 1b,d) is demonstrated in Fig. 2a-c and Supporting Note 9. Specifically, by applying a tensile strain to a magnetized sample we achieve full control over the AF state of $Mn_3Ir$, which is read out by monitoring $H_{PEB}$. The AF state "1" (positive $H_{PEB}$) and AF state "0" (negative $H_{PEB}$) are obtained by down (up) magnetizing the sample, respectively. For demagnetized samples, the application of strain is unable to induce a preferential AF state in $Mn_3Ir$ films, giving rise to zero $H_{PEB}$ (Fig. 2b,c; Supporting Note 5). This result agrees with the common sense that strain (featuring with uniaxial symmetry) alone fails to drive the 180 degrees reversal of the AF order.

The strain stability of the written AF states was tested by exposing sample to different tensile strains without external magnetic field (Fig. 2d). During measurement, the samples were set in the "parallel mode" (the Co magnetization is aligned with the magnetic field to set the AF state, Supporting Note 4). The results show that $H_{PEB}$ of the sample set in the "parallel mode" is insensitive to mechanical deformations (the maximal variation is $< 5\%$), indicating the strain stability of the written AF state. Moreover, we summarize the $H_{EB}$ deviation rate per tensile strain ($\delta|H_{EB}|/\varepsilon_{stretch}$) of reported EB systems (shown in Supporting Note 4), which reveal that our $H_{EB}$ exhibits superior strain stability among other EB systems. In addition to the MOKE measurements reported in Fig. 2d, this strain stability is confirmed also using AHE characterization (Supporting Note 6), emphasizing its great potential to be integrated in advanced flexible electronics to realize strain-invariant magnetic field sensors[45,46].

The robustness of the strain-induced manipulation of the AF state against an external perturbing magnetic field is demonstrated in Fig. 2e,f and Supporting Note 7. Specifically, positive magnetic field pulses were applied to the sample in the AF state "1", while negative magnetic-field pulses were applied to the sample exhibiting AF state "0". After each field pulse, a FM hysteresis loop was measured to probe the state of $Mn_3Ir$. For the sample set in the AF state "1", the magnitude of $H_{PEB}$ varies by about 7% when a perturbing magnetic field of 60 kOe is applied, whereas for the AF state "2", the variation is approximately 10%. Similarly, negative magnetic field pulses applied to the sample in AF state "1" result in a variation of about 3% in $H_{PEB}$ (Supporting Note 7). Overall, the written AF states are quite stable when under large perturbing magnetic fields that are substantially large enough to switch the FM layer of our samples (about 800 Oe).

## Mechanism of setting AF states via piezomagnetism and interfacial Dzyaloshinskii-Moriya interaction

Fig. 3a depicts the initial stages of the time evolution of $H_{PEB}$ of the samples exposed to different tensile strains (see also Supporting Note 10). We vary the stretching step size while increasing the strain from 0 to 1.2% and collect magnetic hysteresis loops to assess the variation of $H_{PEB}$ as a function of both time and strain. It could be inferred from Fig. 3a that the manipulation of the AF state is independent of the stretching step size and magnetizing time, i.e. 120 Oe $H_{PEB}$ can be achieved by a single-step stretching or step-wise increase of strain to 1.2%, and remains stable regardless of magnetizing time. This phenomenon is attributed to the negligible effect of isothermal crystallization and thermal activations[47] on the spin structure of $Mn_3Ir$. By adopting the stretching strategy composed of a single stretching step and accelerating the stretching rate, the AF state could be manipulated within 1 second (Supporting Note 10). The emergence of the faster manipulation of the AF state stands in stark contrast to the isothermal crystallization mechanism, which is responsible for SEB[4,48], where achieving a substantial $H_{PEB}$ necessitates prolonged magnetizing time.

For comparison, the SEB effect governed by isothermal crystallization is demonstrated in Fig. 3b and Supporting Note 12, where the samples are exposed to a tensile strain throughout the entire measuring period of many hours. The time dependence of $H_{PEB}$ is fitted according to the Kolmogorov-Avrami model as $H_{PEB} \propto (1 - 2e^{-(t/\tau)^{\beta}})$, where $\tau$ and $\beta$ are the characteristic relaxation time and exponent of the domain relaxation, respectively[49]. With $\beta$ = 0.035, the characteristic time $\tau$ drops from about 4800 h for $\varepsilon$ = 0.0% to about 680 h at $\varepsilon$ = 1.2% (Supporting Note 12). We also note the exponentially fast rate of setting the $H_{PEB}$ at the very beginning of the process (Fig. 3c): to reach half of the maximal measured value at 1.2% strain, we need only about 30 min.

To understand the rapid onset of $H_{PEB}$, we perform DFT calculations and micromagnetic analysis of the magnetic state of a single grain of $Mn_3Ir$ with the crystallographic axes aligned to the direction of applied strain (Fig. 3d, Supporting Note 11). First, we note shift of the EB loop is opposite in opposite AF domains, see Fig. 1f,g. Thus, in the multidomain film with about equal distribution of AF domains, there will be a finite coercivity, but zero $H_{PEB}$. We stretch the multilayer sample with single-domain-state ferromagnet magnetized by a magnetic field, and this selects the energetically preferred AF domain.

The state of the $Mn_3Ir$ unit cell is determined by the directions of the Mn moments in its three magnetic sublattices, $\mathbf{M}_{1,2,3}$, which are recast into two orthogonal unitless Néel vectors $\mathbf{n}_1 = (\mathbf{M}_2 - \mathbf{M}_1)/(\sqrt{3}M_0)$, $\mathbf{n}_2 = (2\mathbf{M}_3 - \mathbf{M}_1 - \mathbf{M}_2)/(3M_0)$, and the total magnetization $\mathbf{M} = \sum_{i=1}^{3} \mathrm{M}_i$, where $M_0$ is the saturation magnetization of each of magnetic sublattices in the absence of strain (Supporting Note 11). In the ground state $|\mathbf{n}_{1,2}| = 1$. We simulate the experimental situation by applying an IP uniaxial strain parallel to one of the IP hexagonal lattice vectors, here $\hat{\mathbf{x}}$ for concreteness up to $\varepsilon = 5\%$ (Fig. 3d). The mechanical and magnetic degrees of freedom are relaxed and the value of magnetic moments in each of the three sublattices of Mn ions is quantified. We found a linear increase of the sublattice moments $\mathbf{M}_{1,2,3}$ with strain (Fig. 3e). The total moment $\mathbf{M}$ shows two features. There is an almost strain-independent OOP component $M_z \approx 0.02\ \mu_B$ driven by the intrinsic DMI in $Mn_3Ir$ (Fig. 3f), which is also present at zero strain with the sign of $M_z$ set by the selected AF domain. The IP magnetization component perpendicular to the strain direction, $M_y$, reveals piezomagnetism of $Mn_3Ir$. This IP component increases with strain because of a tilt of sublattices due to mechanical distortions (Fig. 3f). The DFT-calculated strain-driven change in $\mathbf{M}_{1,2,3}$ leads to an increase in the Néel vectors (Fig. 3g) as the following equation (Supporting Note 11):

$$n_{1,2}(\varepsilon) = 1 + a\varepsilon \pm \frac{m_y(\varepsilon)}{2} \qquad (1)$$

where $a \approx 1$ for $Mn_3Ir$ is a constant that generally depends on the exchange and magnetomechanical coupling of the particular antiferromagnet, and $m_y = M_y/(3M_0)$ is the IP magnetization normalized. The last term reflects the strain-driven deviation from the ground-state directions of the magnetic sublattices.

A standard phenomenological model of the interfacial exchange coupling proposed by Slonczewski[50,51] considers the coupling between the OOP Co magnetization, $\mathbf{m}_F$, and the uncompensated moment, $M_z$, of $Mn_3Ir$, which is strain insensitive (Fig. 3f), while the in-plane magnetization of piezomagnetic origin is averaged to zero. Thus, this model is not sufficient to explain the experimentally observed features of the exchange bias. We take into account the iDMI[52-54], which couples $\mathbf{m}_F$ with the IP components of the Néel vectors through

$$F_{iDMI} \propto m_F^z(n_{1x} + \sqrt{3}n_{1y} - \sqrt{3}n_{1y} + n_{2y}) \qquad (2)$$

where $F_{iDMI}$ is the energy of the inter-layer coupling. The piezomagnetic change of the Néel vectors (1) leads to an effective increase of iDMI coupling of about 1%. For a general orientation of the $Mn_3Ir$ grain, the piezomagnetic response is reduced upon misalignment between the crystallographic axes and the strain direction, but remains non-zero (Supporting

Note 11). We note in the case of randomly oriented grains the effectively zero net magnetization of piezomagnetic origin does not contribute to the development of PEB because the key role is played by the elongation of magnetic sublattices under tensile strain, but not their mutual tilt. In summary, stretching in a magnetic field selects ferromagnetic domain, the energy barrier between oppositely ordered AF domains is then lowered by about 1% due to iDMI, facilitating switching of AF domains that have higher energy due to unfavorable orientation with respect to the ferromagnetic layer and causing persistent bias after release of the strain when the energy barrier depth is restored preventing further AF switching.

This theoretical estimate of the AF switching barrier between the opposite AF domains by about 1% is comparable with the characteristic change of the energy barrier given by the Arrhenius-Néel law of about 4% based on the attempt frequency estimates in $Mn_3Ir$[4] and characteristic time $\tau$ extracted from the Kolmogorov-Avrami model see Fig. 3h and Supporting Note 12. An additional decrease of the energy barrier can be related to further changes of the exchange and anisotropy energies in $Mn_3Ir$ and $[Co/Pt]_3$ layers as well as the Co magnetization in the film that forces easier depinning of AF domain walls in $Mn_3Ir$. Attributed to the reduced energy barrier through the strain-mediated iDMI effect, the AF states can be established as soon as mechanical strain is applied. As shown in Supporting Note 12, our approach overcomes the speed limitations imposed by conventional isothermal methods based on isothermal crystallization mechanism. We envision that our approach can be extended to AF/FM systems prepared on piezoelectric substrates enabling much faster manipulation of AF states through electrical field[27].

## Discussion and outlooks

In practical memory applications, the ability to extend a memory cell into memory arrays have to be evaluated. We propose two possible schemes for multi-cell writing process, one of which is demonstrated. As is shown in Fig. 4a, we extend the single memory cell to an example array of 16 memory units, the memory cells are read out by measuring the AHE signals. Here, we demonstrate a two-step multi-cell write scheme, with the cells 1, 2, 3, 4 written into the binary information of "1", "0", "1", "0", respectively. To do that, we first polarize the FM layer of each cell with predefined magnetic field (i.e., with cells 1 and 3 magnetized downward, cells 2 and 4 magnetized upward). Afterwards, the substrate is stretched to $\varepsilon = 1.2\%$ and then released to mechanically relaxed state, this allows to switch the AF layer into the desired moment direction. The viability of this scheme is verified by reading out the status of the written AFM cells. As is shown in Fig. 4b, the AHE curves of cells 1-4 indeed represent the binary information of "1", "0", "1", "0", respectively. The binary information is also decoded by the $H_{PEB}$ indicated in the AHE curves. Our results thus demonstrate a simultaneous and deterministic manipulation of different AF states within the array through a magnetization-stretch operation, which avoided the collective switching of the memory cells. Furthermore, by replacing the permanent magnets with magnetic write heads, the AF bit size can be reduced to the nanoscale, which satisfies the size requirements for AF tunneling magnetoresistence (TMR) arrays. Alternatively, considering a rather high remanent magnetization of the FM layer, the formation of a preferential AF state can be readily achieved even in the absence of an external magnetic field, provided that the $[Co/Pt]_3$ multilayers have been magnetized prior to stretching (Supporting Note 8). This opens an appealing possibility to use SOT-based switching of the FM layer to manipulate the AF state all-electrically without the need for magnetic fields[55]. In another possible scheme, the AF cells could be written by overall magnetization followed by localized strain excitation, although at present it faces critical challenges of applying a large yet localized strain to obtain a sizable effect. This also drives further extrapolation of new AF systems that exhibit higher mechanical switching efficiency. Furthermore, as our approach is also valid in systems featuring with in-plane (IP) magnetic anisotropy (Supporting Note 13), the piezomagnetic manipulation of the AF state in AF/FM stacks presents a novel method to introduce tailorable exchange bias effect in flexible magnetic field sensors, expanding applicability of our approach to smart wearable electronics. For example, arranging mechanically flexible magnetic field sensors in Wheatstone bridges (a specific circuit architecture used for high-precision sensing) offers major enhancement in their sensitivity and thermal stability yet is rather challenging in fabrication relying on established methods due to the necessity of having opposite exchange bias fields in the neighboring AF/FM-based spin valve units[56]. As our approach does not rely on high processing temperatures, it is particularly well suited for the realization of Wheatstone bridges of flexible magnetic sensors, which not only simplifies the fabrication process and enhances production efficiency, but also prevents the generation of Joule heating (Supporting Note 13).

## Conclusion

In conclusion, we report an isothermally nonvolatile and deterministic switching of AF states in triangular $Mn_3Ir$ thin films in proximity to a FM $[Co/Pt]_3$ multilayer. DFT and micromagnetic analysis reveal that the underlying mechanisms are piezomagnetism of $Mn_3Ir$ and the iDMI effect at the AF/FM interface. These couplings effectively reduce the energy barrier for $Mn_3Ir$ domain switching by about 4%, thereby overcoming the speed limitation of conventional isothermal approaches relying on isothermal crystallization. Furthermore, the strain-manipulated AF states exhibit excellent robustness against external perturbations such as magnetic field and external strain. Our approach is valid in both OOP and IP systems, thus facilitating the development of interactive flexible electronics and energy-efficient isothermal spintronic memories.

## Methods

### Fabrication process

Multilayers with perpendicular magnetic anisotropy (PMA) consist of Ta(4)/Pt(2)/$Mn_3Ir$(6)/[Co(0.6)/Pt(1)]$_3$/Pt(1) stacks (in brackets, the thicknesses of individual layers is indicated in nm). In the stacks, Ta/Pt layers served as a buffer layer to diminish a surface roughness of the polymeric foil and favor the growth of $Mn_3Ir$ layer with the (1 1 1) texture. The AF layer $Mn_3Ir$ (6 nm) was sufficiently thick to induce a pronounced interfacial coupling effect. [Co/Pt]$_3$ multilayer stack is a FM layer. A capping layer of Pt (2 nm) was deposited to prevent the multilayers from oxidation. These samples were deposited on 50 $\mu$m thick PI foils (Wuxi Baike Electronic Materials, China) at room temperature using a DC magnetron sputtering (Adnano-tek, China) in an Ar atmosphere (Ar pressure: $3 \times 10^{-3}$ mbar) in a chamber with a base pressure of less than $5 \times 10^{-7}$ mbar. The dimensions of the PI foils are $10 \times 30$ mm$^2$, and the metal multilayers were grown in a shape of a circle with the diameter of 6 mm using a mask template. The absence of an external magnetic field during deposition induces zero perpendicular exchange bias field ($H_{PEB}$) in the as-deposited samples. The thickness of each layer was determined through the control of the deposition time at the measured deposition rate of 0.034 nm/s (50 W) for Ta, 0.062 nm/s (50 W) for Pt, 0.053 nm/s (80 W) for $Mn_3Ir$, and 0.032 nm/s (50 W) for Co.

An array of units for the AHE characterization was fabricated through lift-off processes and sputtering deposition. A photoresist layer was spin coated on the PI substrate, then the electrodes and circuits were patterned at room temperature through UV lithography (MA6, SUSS MicroTec, Germany). Following the lithography, the electrode layer composed of Pt (50 nm) was deposited via magnetron sputtering. The multilayer stacks with PMA consisting of Ta(4)/Pt(2)/$Mn_3Ir$(6)/[Co(0.6)/Pt(1)]$_3$/Pt(1) stacks (in brackets, the thicknesses of individual layers is indicated in nm) were patterned into four units and deposited through the same procedures.

## Structural characterization

Cross-sectional TEM images were captured using a cold-field TEM (CFTEM, F200, JEOL, Japan). The images were used to characterize the crystal structure and texture of the samples. The TEM samples were prepared through a Focused Ion Beam (FIB, Auriga, Germany) etching. The crystal quality was estimated by X-Ray Diffraction (XRD, D8 Discover, Bruker, Germany) and Small Angle X-Ray Scattering (SAXS, Xeuss 3.0, UHR, France) with Cu $K_\alpha$ ($\lambda$ = 1.541 Å) radiation.

**SAXS Measurement**. For each SAXS pattern, the exposure time was 10 min. The 2D scattering patterns are ring-shaped. Therefore, 1D diffraction curves were obtained by averaging the scattering intensity within the scattering plane shown in 2D SAXS patterns. The averaging equation to obtain the 1D diffraction curves is shown below:

$$I_{av}(q) = \frac{\int_{\phi_2}^{\phi_1} I_\phi(q)\mathrm{d}\phi}{|\phi_2 - \phi_1|} \tag{3}$$

In equation (3), $q$ is the scattering vector, $I$ represents the scattering intensity, $\phi_1$ and $\phi_2$ denote the central angle within the scattering plane, $\phi$ is defined as the central angle between the chosen direction to the stretching direction. The relationship between the diffraction angle $2\theta$ and $q$ is given by the equation:

$$q = \frac{4\pi\sin\theta}{\lambda} \tag{4}$$

The geometry of the 2D scattering pattern is shown in Supporting Note 1. According to the equations (3) and (4), the OOP 1D diffraction curves were calculated by selecting the entire scattering plane ($\phi_1 = 0°$, $\phi_2 = 360°$). 1D diffraction curves parallel to the scattered wave vector were calculated by selecting the central angle range of $\phi_1 = 5°$ and $\phi_2 = 5°$.

**Peak Fitting**. The XSACT software is used to search for approximate positions of the certain texture through Lorentz fitting. The positions are input into the ORIGIN software to fit the peaks more accurately. The fitting is done to the Lorentz function. This peak fitting method is available for both XRD curves and 1D SAXS curves.

**Calculation of Crystal Parameter**. Only the $Mn_3Ir$ texture is visible in XRD and SAXS patterns, owing to the polycrystalline states and thin profiles of other layers. By taking the crystal structure of $Mn_3Ir$ (FCC) into consideration, the crystal parameter of the $Mn_3Ir$ layer ($d_c$) was calculated as follows:

$$d_{\mathrm{c}} = \frac{\lambda\sqrt{a^2 + b^2 + c^2}}{2\sin\theta} \tag{5}$$

In equation (3), the parameters $a$, $b$ and $c$ represent indices of the crystal plane, $\lambda$ denotes the wavelength of the X-ray beam ($\lambda = 1.541$ Å), $\theta$ is half of the diffraction angle or scattering angle of the peak.

### Magnetic measurements

The magnetic properties of the samples were measured using a Superconducting Quantum Interference Device Vibrating Sample Magnetometer (SQUID-VSM, Quantum Design, USA), a differential polar-MOKE magnetometry system with commercial microscope (MOKE, Evico Magnetics, Germany), and a MOKE Microscopy (KMPL-Spin-X, Truth Instruments, China). The Anomalous Hall Effect (AHE) measurements were carried out using a custom-built magnetoresistive system composed of a Tensormeter setup (HZDR Innovation GmbH, Germany), and a Physical Property Measurement System-Magnetoresistance Measurement (PPMS-MM, Quantum Design, USA).

**SQUID**. The magnetic anisotropy was estimated by measuring IP and OOP hysteresis loops. Before the measurement, the deposited samples (6 mm diameter circle) were cut into $3 \times 3$ mm$^2$ pieces to fit the sample holder of the SQUID magnetometer. The dimensions of the cut sample were measured using a commercial micrometer. The cut-out area was located at the center of the original sample. The IP and OOP hysteresis loops were obtained by fixing the sample plane parallel and perpendicular to the magnetic field, respectively. The measuring temperature was fixed at room temperature (300 K) and the maximum magnetic field was large enough to saturate the sample. The hysteresis loops of the samples were obtained after subtracting the paramagnetic signal.

**MOKE Magnetometry and Microscopy**. OOP hysteresis loops were obtained by a MOKE magnetometry at room temperature under an OOP magnetic field generated through electromagnets powered by a quadrupolar power supply. The reported normalized magnetization corresponds to the normalized Kerr signal. From the normalized hysteresis loops we calculated $H_C$ and $H_{PEB}$. The magnetic domain patterns were recorded through a MOKE microscopy at room temperature under an OOP magnetic field generated through electromagnets powered by a dipolar power supply. During the measurement, the applied magnetic field was varied and a magnetic domain pattern was automatically captured.

**Magnetotransport**. The AHE curves of the sample composed of [PI foil]//Ta(2)/Pt(4)/$Mn_3Ir$(6)/[Co(0.6)/Pt(1)]$_3$/Pt(1) (numbers in brackets are given in nm) were recorded through a custom-built magnetoresistive system. We used four-point probe method to measure the AHE of the sample, and the applied magnetic field was perpendicular to the film surface. The detailed configuration of the measurement and the method to remove the offset can be referred to Supporting Note 6.

**Magnetoresistance**. For the measurement for each unit, magnetoresistance (MR) curves were measured via a four-point method with the measuring system consisting of a current source (B2912B, Keysight), a nanovoltmeter (2182A, Keithley) and a pair of electromagnets. During the measurement, the current $J$ generated by the current source and the magnetic field produced by the electromagnets were applied parallel to the pinning direction if not specified otherwise. The full GMR loop was measured by applying magnetic field large enough to fully switch the magnetic moment of the reference layer. The minor GMR loop was measured by applying magnetic field sufficient to switch the magnetic moment of the free layer but not strong enough to switch the reference layer. The voltage $U$ across the sample was measured using a nanovoltmeter. The resistance $R$ of the sample was calculated as $R = U/I$. The MR was calculated as:

$$\mathrm{MR} = \frac{R - R_{\mathrm{min}}}{R_{\mathrm{min}}} \tag{6}$$

$R_{\mathrm{min}}$ was obtained after subtracting the linear background.

## Strain application

For the samples deposited on PI foils with the dimensions of $10 \times 30$ mm$^2$, the tensile strain was applied through a custom-built tensile stretcher. By fixing the PI foils using grippers at the initial distance $d_0$ of 10 mm, the tensile strain was applied by stretching the sample to a distance $d$. The strain applied on the PI foil was calculated as $\varepsilon_{\mathrm{PI}} = (d - d_0)/d_0$. The strain was applied at the stretching speed of 5 $\mu$m/s if not noted otherwise.

## DFT simulations

For DFT calculations of $Mn_3Ir$, we employ the Vienna *ab initio* simulation package (VASP)[57], using the generalized gradient approximation with the Perdew-Burke-Ernzerhof (PBE) functional[58]. We include spin-orbit coupling (SOC) self-consistently in all of our calculations, and initialize the Mn magnetic moments in their noncollinear antiferromagnetic ground-state order, with IP triplets of Mn moments rotated 120 degrees relative to each other. We use the Projector-augmented-wave (PAW) method[59] using the standard VASP PAW pseudopotentials, and take Ir : $5d^8 6s^1$ and Mn : $3p^6 3d^6 4s^1$ electrons as valence. We use a Gamma-centered 10 × 10 × 8 reciprocal mesh to sample the Brillouin zone and an 800 eV kinetic energy cutoff for our plane-wave basis set. We then allow atomic positions and the lengths of the unit cell orthogonal to the direction of applied strain to relax until forces on atoms are again less than 0.01 eV/Å, and the stresses are below a stopping criterion which is internally scaled in VASP based on the selected force value. The relaxed lattice parameters for the hexagonal unit cell (which we use rather than the cubic cell for convenience, since in the hexagonal basis the Mn moments lie perfectly parallel to the a-b plane) are $a$ = 5.25 Å and $c$ = 6.44 Å.

In the calculations of $Mn_3Ir$'s piezomagnetic response, for a given percent of uniaxial strain we fix one IP lattice constant to the desired, strained length. We then allow atomic positions and the lengths of the unit cell orthogonal to the direction of applied strain to relax until forces on atoms are again less than 0.01 eV/Å. Finally, we calculate the magnetization density projected onto the Mn atoms using the relaxed, strained structures.

**Phenomenological description of the temporal evolution of exchange bias**

The magnetic state of $Mn_3Ir$ was described based on the so-called spin frame[60]. The unit cell is characterized by three magnetic vectors: two Néel vectors $\mathbf{n}_{1,2}$ that are orthogonal in the ground state, and magnetization. Based on the magnetic symmetry of $Mn_3Ir$, we derive the expression for the phenomenological magnetic energy that includes the uniform exchange, anisotropy, and Dzyaloshinskii-Moriya interaction (DMI). In particular, the latter leads to the development of a finite OOP magnetization along $3_z$ axis. The coupling between $Mn_3Ir$ and perpendicularly magnetized Co can be described via the interfacial DMI[50,53,54] that depends on the direction of magnetization of Co and the direction of $\mathbf{n}_{1,2}$, but not the uncompensated magnetization, see Supporting Note 11 for details.

A finite strain in $Mn_3Ir$ strongly alters the magnetic subsystem. This includes modifications of the exchange interactions and appearance of a finite magnetization of the piezomagnetic origin. The change of the uniform exchange is reflected in the elongation of magnetic moments and, respectively, the Néel vectors. An enhancement of the interfacial DMI coupling due to the change of $\mathbf{n}_{1,2}$ alters energy barriers for the spontaneous relaxation of the magnetic system, that can be described by the model proposed in Ref. Xi *et al.*[49]. The change of the energy barrier calculated based on the exchange bias field measured experimentally is matches the theoretical prediction fairly well (Supporting Note 12).

## Data availability

All data that support the plots within this paper and other findings of this study are available from the corresponding authors upon reasonable request. Source data are provided with this paper.

## Acknowledgements

This work was supported by the National Natural Science Foundation of China (U24A6001, 52522108, U23A20551, 12474126, 52371205, 52127803), Zhejiang Provincial Natural Science Foundation of China (LD24E010001), the Ningbo Key Scientific and Technological Project (2024Z147), the European Commission HORIZON RIA (project REGO; ID: 101070066) and ERC grant 3DmultiFerro (project number: 101141331). Computational resources for the DFT calculations were provided by ETH Zürich's EULER cluster (project ID: eth3).

## Author contributions

X. Bao, O.V. Pylypovskyi, H. Yang, Y. Xie, D. Makarov and R.-W. Li conceived the project idea. X. Bao, H. Yang, Y. Xie, J. Wang, D. Makarov and R.-W. Li designed the experiments. H. Yang, Y. Xie, D. Makarov and R.-W. Li supervised the project. X. Bao fabricated the materials and conducted the measurements. O.V. Pylypovskyi conducted micromagnetic analysis. S.F. Weber and N.A. Spaldin performed DFT calculations and respective data analysis. D. Faurie, F. Zighem and X. Wang provided the data and analysis of strain distributions. P. Makushko, L. Guo, P.T. Das, X. Hong contributed to the electromagnetic measurements. J. Liang, R. Zou, H. Jiang and D. Han contributed to the fabrication of Wheatstone bridge. X. Bao, O.V. Pylypovskyi, H. Yang, N.A. Spaldin, D. Makarov and R.-W. Li wrote and revised the manuscript with inputs from all authors. All authors discussed the results and provided comments on the manuscript.

## Competing interests

The authors declare no competing interests.

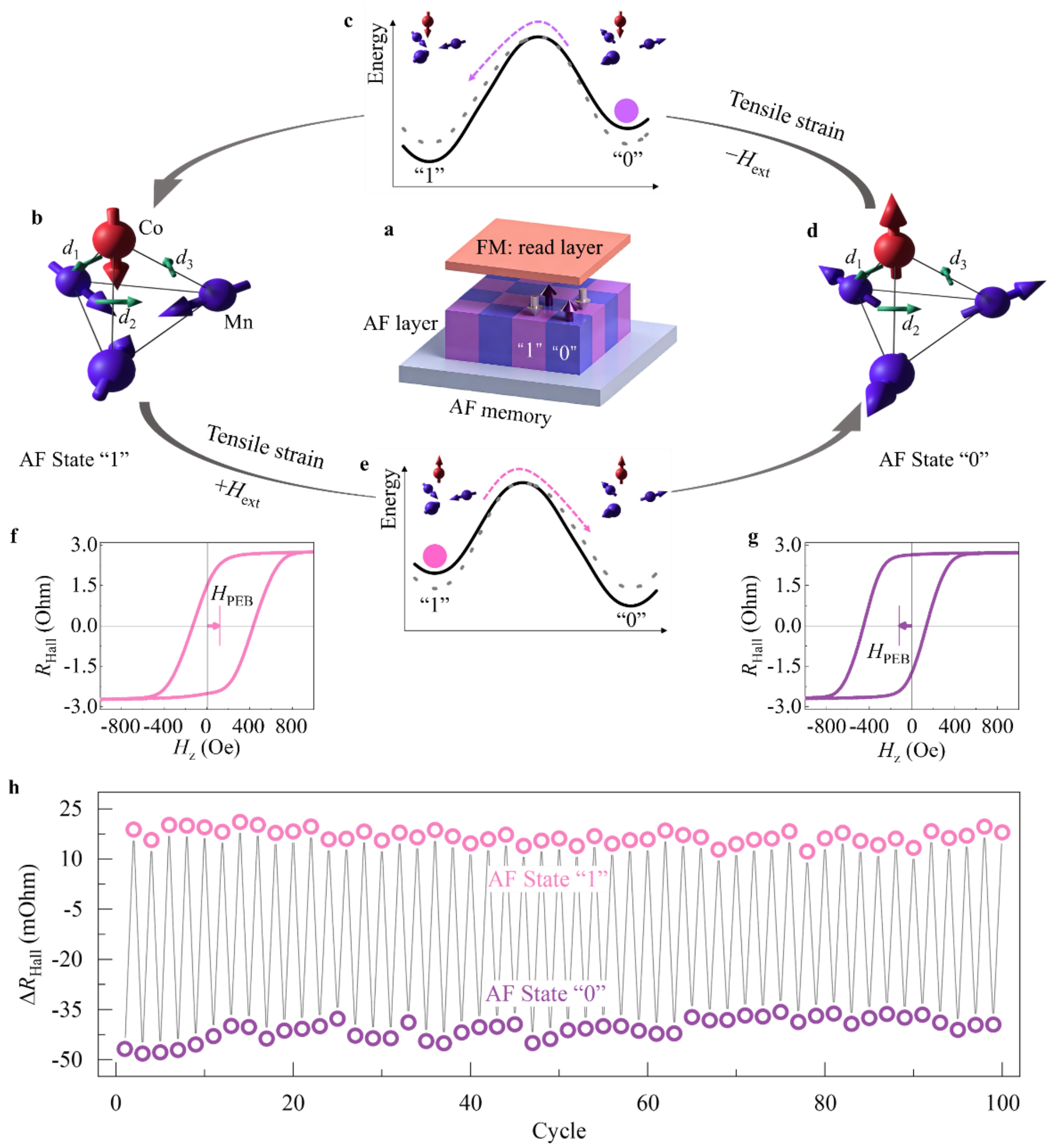

**Fig. 1| Strain-induced switching of the antiferromagnetic state: towards AF-based memories. a**, Schematic illustration of an AF memory device composed of an AF layer and a FM multilayer. The FM multilayer is utilized to read out the magnetic state of the AF layer. **b**, The equilibrium configuration of AF state 1 (bit "1") with inward Mn spins (violet) and downward Co spin (red). Here, green arrows show DMI coupling between Mn and Co spins. **c**, The schematic illustration of energy levels between two AF states during the switching from AF state "0" to AF state "1". This switching is achieved under the application of a negative magnetic field $-H_{ext}$ and tensile strain. The initial energy symmetry between AF domains (grey dashed curve) is broken, resulting in energy asymmetry between AF domains (black solid curve). **d**, The equilibrium configuration of AF state 2 (bit "0") with outward Mn spins and upward Co spin. **e**, The schematic illustration of energy levels between two AF states during the switching from AF state "1" to its opposite AF domain (AF state "0"). This switching is achieved under the application of a positive magnetic field $+H_{ext}$ and tensile strain. **f**, The

hysteresis loops revealing AF state "1" and **g**, AF state "0", these loops are measured via AHE. **h**, The switching behavior between the two AF states over 100 write-read cycles. The magnetic state is read out at remanence, i.e. after removal of the mechanical strain and external magnetic field. During the switching process, the magnitude of applied magnetic field and mechanical strain is 2 kOe and 1.2%, respectively.

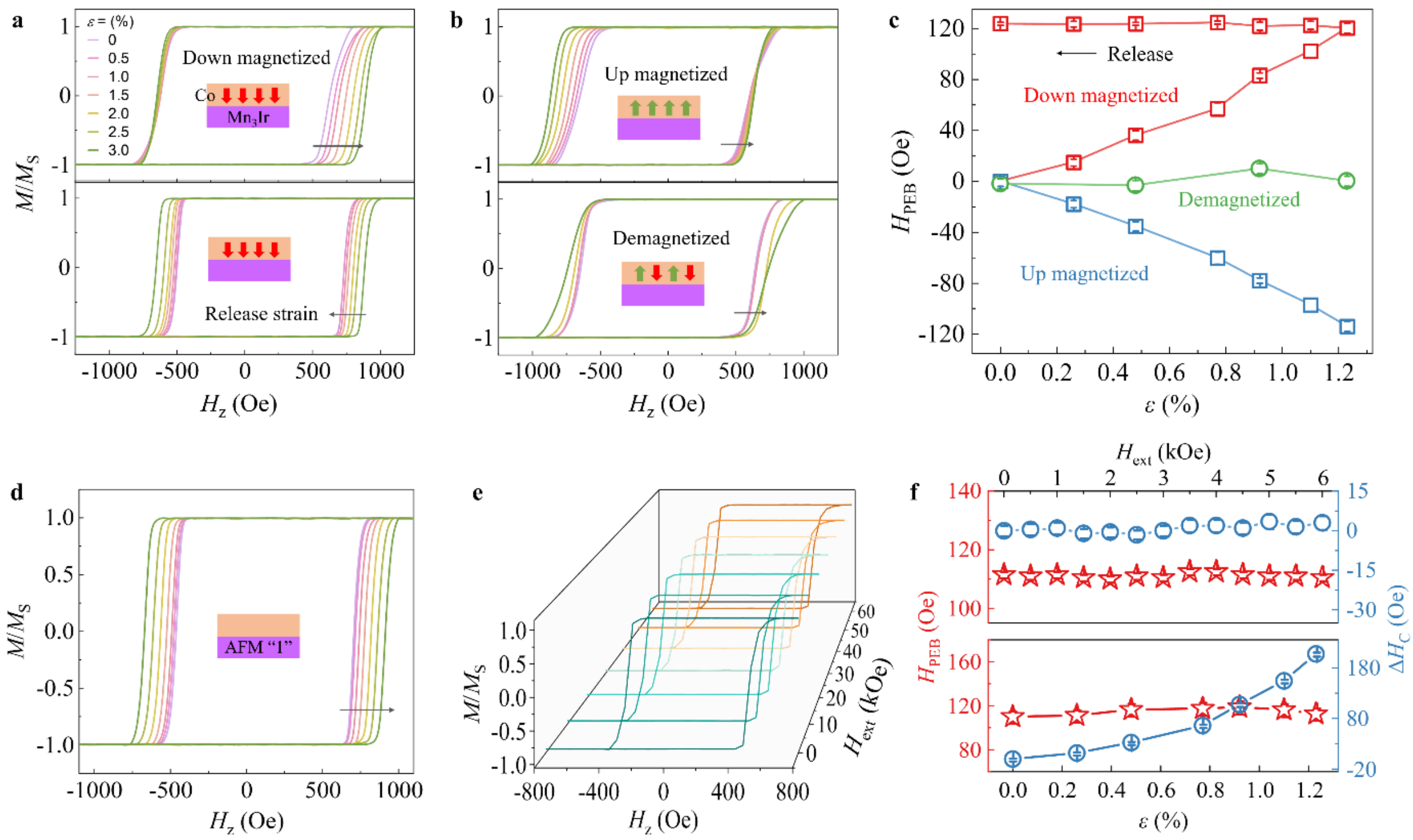

**Fig. 2| Deterministic manipulation of nonvolatile antiferromagnetic states. a**, Normalized OOP Kerr loops with down magnetized state upon application (upper panel) and release of strain (lower panel). **b**, Normalized OOP Kerr loops with up magnetized state (upper panel) and demagnetized state (lower panel) at different $\varepsilon$. The magnitude of the magnetizing field is 2 kOe. **c**, Strain dependence of the measured $H_{PEB}$ with the sample in three different magnetized states. The data were calculated from the hysteresis loops shown in panels (a) and (b). **d**, Normalized OOP MOKE loops of the mechanically relaxed sample set in the AF state "1" under different $\varepsilon$. It should be noted that during this measurement, no magnetic field was applied during stretching process and the sample was left in "parallel mode". Details on the process are in Supporting Note 4. **e**, Normalized hysteresis loops of the mechanically relaxed samples exposed to an external perturbing magnetic field ($H_{ext}$) ranging from 0 to 60 kOe. **f**, The dependence of $H_{PEB}$ and $\Delta H_C$ with strain and the perturbing magnetic field for samples set in the AF state "1".

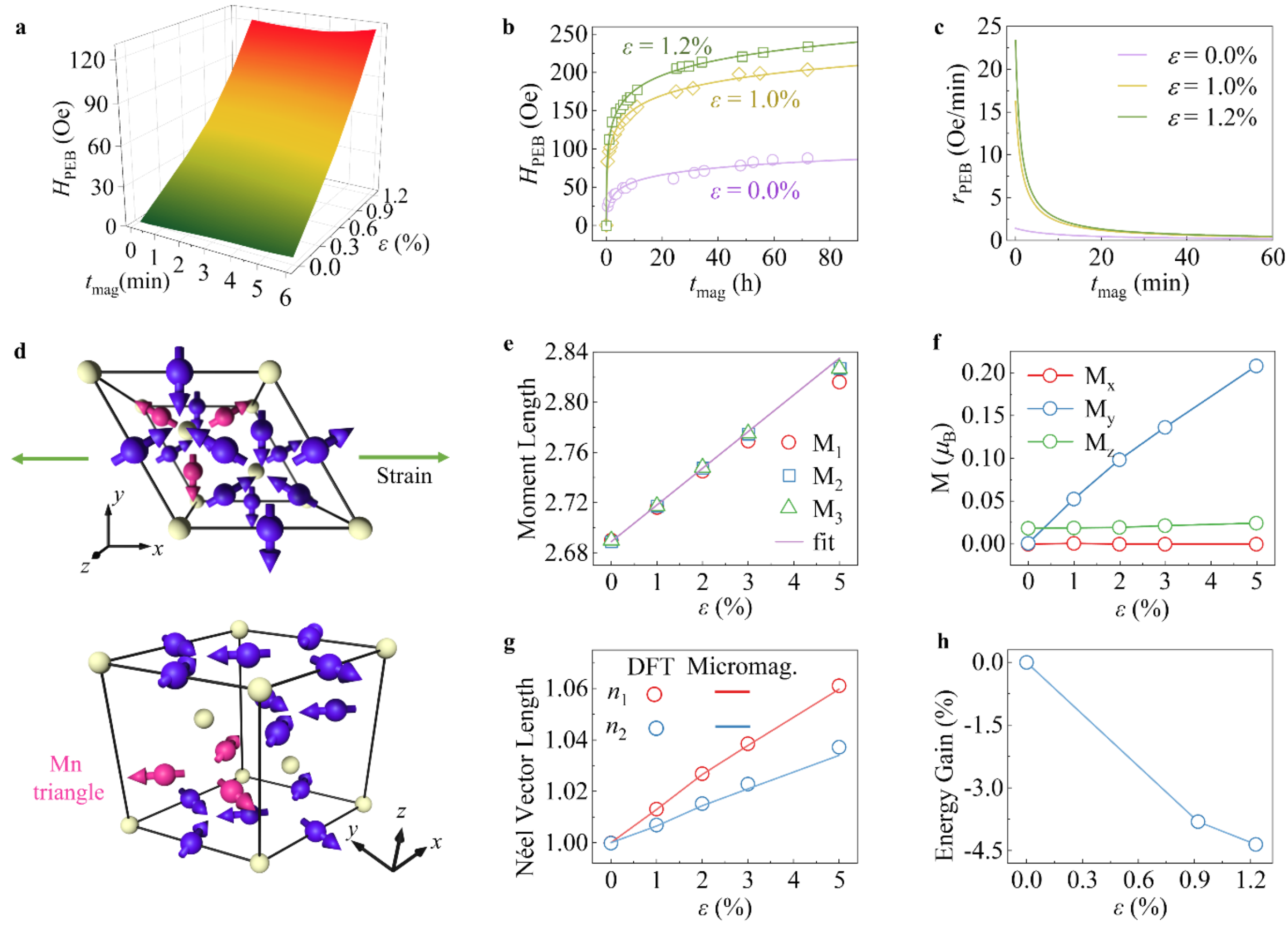

**Fig. 3| Fundamentals of the strain-induced manipulation of the AF state in $Mn_3Ir$. a**, Initial stages of the time evolution of $H_{PEB}$ in the samples exposed to different tensile strain. The color code represents the strength of $H_{PEB}$. The strain is increased step-wise to a specified value using different step sizes, details of the measuring protocols are provided in Supporting Note 10. The measurement is done with an increment of 1 min. During the entire stretching process, the sample is exposed to magnetic field of 2 kOe prior to the strain. **b**, Magnetizing time, $t_{mag}$, dependence of $H_{PEB}$ for the samples exposed to different tensile strains. Symbols represent experimental data measured under different $\varepsilon$. Solid curves delineate the fitted results through the Kolmogorov-Avrami model. In this measurement, strain is applied to the sample prior to the magnetizing process; details of the measuring protocols are provided in Supporting Note 12. **c**, Rate of increase of $H_{PEB}$, $r_{PEB}$, with magnetizing time, the data are calculated from panel (b). **d**, Schematics illustrating the magnetic structure of $Mn_3Ir$. The direction of uniaxial strain is indicated. **e**, Calculated $\Delta m$ of three magnetic moments in the unit cell as a function of tensile strains using DFT. A linear fit to the calculated values (symbols) is shown with the solid line. **f**, DFT-calculated components of the total magnetization per unit cell as a function of strain. **g**, Variation of the Neél vectors $n_{1,2} = |\mathbf{n}_{1,2}|$ with the applied strain $\varepsilon$ calculated using DFT and micromagnetic analysis. **h**, Calculated change in the energy barrier as a function of strain for the reversal of $Mn_3Ir$ magnetic domains.

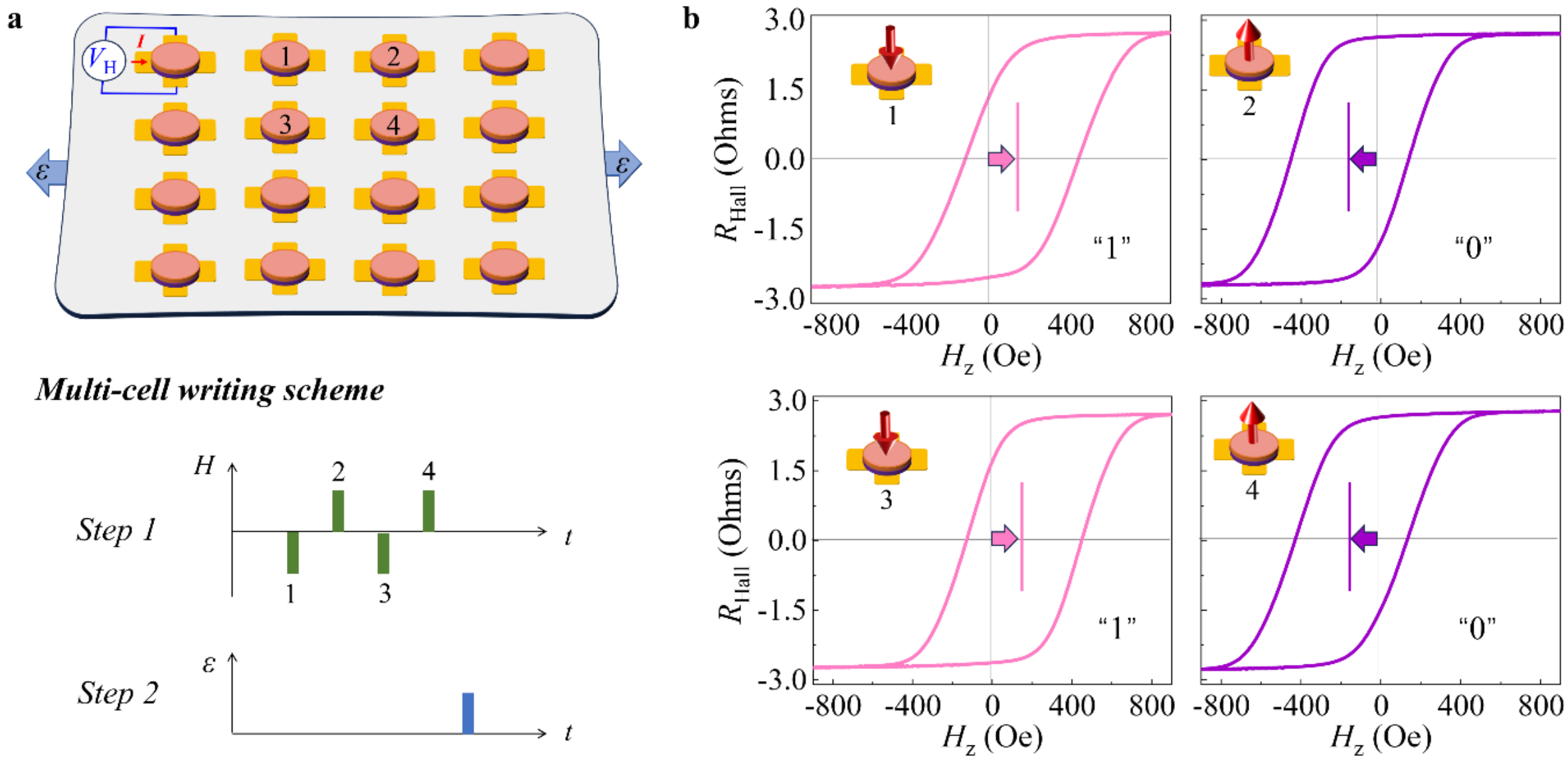

**Fig. 4| Demonstration of multi-cell operation in AF memory array. a**, Schematic figure of the AFM memory arrays and the multi-cell (represented by 1, 2, 3, 4, respectively) write scheme. Each memory cell is read by AHE measurement. **b**, The AHE curves for the four written unit cells shown in panel (a). Cells 1-4 represent the binary information of "1", "0", "1", "0", respectively. The binary information is also decoded by the $H_{PEB}$ indicated by the horizontal arrows in the AHE curves.